\documentclass[aps,prx,superscriptaddress,notitlepage,twocolumn]{revtex4-1}
\usepackage{graphicx,enumerate,verbatim,bbold}
\usepackage{amsmath,amssymb,amsthm,float,mathrsfs}
\usepackage{dsfont}
\usepackage[colorlinks]{hyperref}
\usepackage[T1]{fontenc}    
\usepackage[utf8]{inputenc}
\usepackage{lmodern}
\usepackage{braket}
\usepackage{multirow}
\usepackage{bm}
\usepackage[caption=false]{subfig}
\usepackage{dcolumn}
\usepackage{hyperref}

\usepackage[mathlines]{lineno}
\modulolinenumbers[5]
\usepackage[usenames,dvipsnames]{color}
\newcommand{\beginsupplement}{%
        \setcounter{table}{0}
        \renewcommand{\thetable}{S\arabic{table}}%
        \setcounter{figure}{0}
        \renewcommand{\thefigure}{S\arabic{figure}}%
        \setcounter{equation}{0}
        \renewcommand{\theequation}{S\arabic{equation}}%
     }

\begin{document}

\title{Scalable and robust quantum computing on qubit arrays with fixed coupling}
\author{N.H. Le}

\affiliation{Advanced Technology Institute and Department of Physics, University of Surrey, Guildford GU2 7XH, United Kingdom}

\author{M. Cykiert}
\affiliation{Advanced Technology Institute and Department of Physics, University of Surrey, Guildford GU2 7XH, United Kingdom}

\author{E. Ginossar}
\affiliation{Advanced Technology Institute and Department of Physics, University of Surrey, Guildford GU2 7XH, United Kingdom}

%% ****** Abstract of paper ****** %
\begin{abstract}
We propose a scheme for scalable and robust quantum computing on two-dimensional arrays of qubits with fixed longitudinal coupling. This opens the possibility for bypassing the device complexity associated with tunable couplers required in conventional quantum computing hardware.  Our approach is based on driving a subarray of qubits such that the total multi-qubit Hamiltonian can be decomposed into a sum of commuting few-qubit blocks, and then efficient optimization of the unitary evolution within each block. Each driving pulse can implement a target gate on the driven qubits, and at the same time implement identity gates on the neighbouring undriven qubits, cancelling any unwanted evolution due to the constant qubit-qubit interaction. We show that it is possible to realise a universal set of quantum gates with high fidelity on the basis blocks, and by shifting the driving pattern  one can realise an arbitrary quantum circuit on the array. Allowing for imperfect Hamiltonian characterisation, we use robust optimal control to obtain fidelities around 99.99\%  despite $1\%$ uncertainty in the qubit-qubit and drive-qubit couplings, and a detuning uncertainty at 0.1\% of the qubit-qubit coupling strength. This robust feature is crucial for scaling up as parameter uncertainty is significant in large devices.
\end{abstract} 
\maketitle

\section*{Introduction}
Great progress has been achieved recently in various physical platforms for quantum computing, most notably is the 54-qubit programmable superconducting processor \cite{Arute2019}. High fidelity two-qubit gates were also demonstrated for trapped ions \cite{Ballance2016}, neutral atoms in optical tweezers \cite{Levine2019}, and spin qubits in silicon \cite{He2019,Huang2019} and GaAs \cite{Nichol2017}. These experimental implementations are based on tunable coupling between qubits where the interaction is switched on only when the two qubit gates are implemented. In solid state quantum computers, tunable couplers typically involve more circuit elements and require their own external control for tuning the magnitude of the interaction \cite{Arute2019,Chen2014,He2019} leading to  overheads in fabrication and wiring. For solving real-world problems, a quantum computer  needs a large number of qubits \cite{Fowler2012},  and the complexity of the tunable couplers adds to the technological difficulties in scaling up the device. In contrast, fixed couplers do not require the extra components for controlling the magnitude of the interaction, resulting in a substantial simplification of the hardware architecture and hence a significant advantage for scaling up. 

\begin{figure*}[t]
\centering
\includegraphics[width=2\columnwidth]{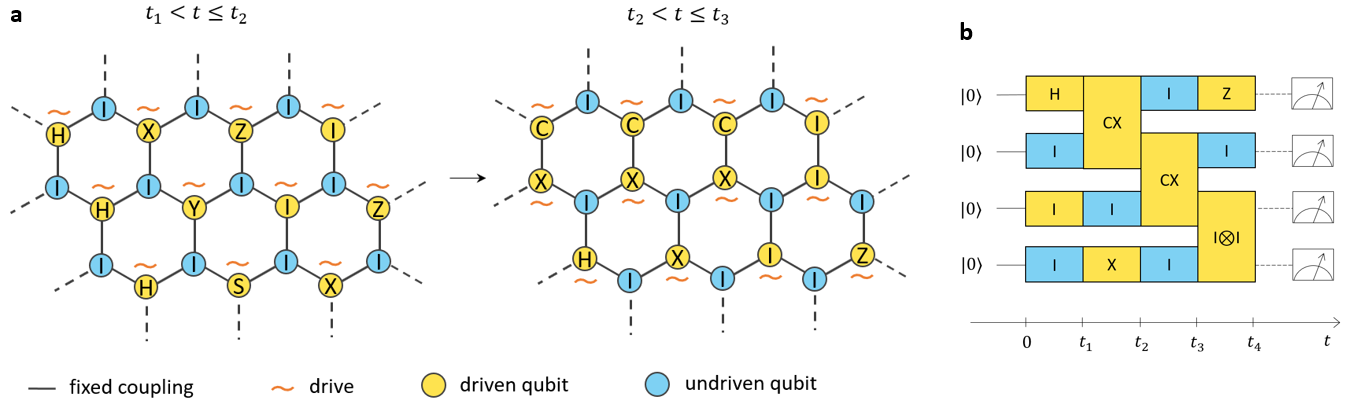} 
\caption{\label{fig:circuit} a) An example of implementing quantum gates on an array of qubits with fixed couplings. In any given step only a sub-array can be driven. This sub-array can be chosen to satisfy specific requirements so that the array's Hamiltonian can be decomposed into commuting few-qubit blocks. Each drive can implement a gate on the driven qubit, and through its combination with the fixed qubit-qubit couplings also implements an identity operator on the neighboring undriven qubits. In the next step a different subarray is driven for implementing gates on a different set of qubits. Here C-X on two adjacent qubits denotes the CNOT gate and $I$ the identity gate. b) Illustration of a  quantum circuit in our scheme. The key feature is that any idle interval between gates can be used to realize with an active identity gate, for preventing unwanted evolution due to the fixed couplings. }
\end{figure*}

An important requirement for quantum computing with fixed coupling is the ability to cancel the unwanted evolution due to the fixed interaction on qubits where no gate is needed. In NMR quantum computing, where the qubits have fixed longitudinal couplings, this is achieved by applying a series of cleverly designed of refocusing pulses \cite{Vandersypen2001,Tsunoda2020}. For large arrays of qubits these series become increasingly complex, which is a bottleneck for scaling up \cite{Tsunoda2020}. In this paper we describe a simple method for quantum computing on qubit arrays with fixed coupling without refocusing pulses. Instead, we rely on a key observation that, by driving a specific subarray, one can implement any gate on the driven qubits, and \emph{at the same time} implement an identity operator on all undriven qubits, effectively cancelling the unwanted evolution on these undriven qubits. Any arbitrary quantum circuit can then be implemented by changing the driven subarray between the time steps. An illustration of the driving pattern and the implementation of gates is given in Fig.~\ref{fig:circuit}. Our method can be scaled up to an arbitrarily large array in a straightforward manner, opening an alternative pathway for a simplified quantum computer hardware architecture based entirely on fixed coupling.

In principle, designing the subarray could be difficult. This is because simulating a constantly interacting system of qubits is in general not possible due to the exponential wall: the cost in memory and time increases exponentially with the number of qubits. Thus, one cannot predict the unitary gate implemented by a driving pulse. In our method this problem is avoided, because the driven subarray can be chosen such that the total Hamiltonian of the system can be decomposed into a sum of commuting blocks of only a few qubits. Each block has a low dimensional Hilbert space, and thus its unitary evolution can be simulated and optimized efficiently. This decomposition exists when the qubit-qubit coupling term is longitudinal, i.e., diagonal in the computational basis, for example, the ZZ interaction. 

An appealing feature of our method is the robustness of the gates against uncertainty in all the physical parameters of the array. By using robust optimal control we find pulses for realizing gates with fidelities around $99.99\%$ despite a $1\%$ uncertainty in all the qubit-qubit and drive-qubit couplings, and a detuning uncertainty at 0.1\% of the qubit-qubit coupling strength. This robustness of the fidelity against uncertainties is crucial for an architecture with entirely fixed couplers because it is not possible to isolate a qubit or qubit pair for a precise measurement of the parameter values, and hence there is always a significant residual uncertainty even after the device characterisation process. 

This paper is organised as follows: We first describe the key details of our method, including the driving pattern to make the Hamiltonian decomposable, the application of a universal set of gates, and the implementation of an arbitrary circuit. Next, we show how to use optimal control to make the gates robust against parameter uncertainty in the Hamiltonian. Finally, we discuss potential physical realisations of our method.   

\section*{Results}

We first describe our method for implementing an arbitrary quantum circuit on qubit arrays with fixed longitudinal coupling. We consider a system of qubits coupled by fixed nearest-neighbor longitudinal interaction, i.e., an interaction that commutes with the bare qubit's Hamiltonian. For simplicity we choose the ZZ interaction, which has been realised  experimentally for superconducting qubits \cite{Johnson2011,Collodo2020}. When a subset of qubits is driven by external fields, the system's Hamiltonian is
\begin{align}
    \mathcal{H}(t)=&-\sum_{j}\frac{\omega_j}{2}\sigma_j^z+\sum_{j\in \mathcal{L}} d_j E_j(t) \sigma^x_j+\sum_{jk}J_{jk}\sigma_j^z\sigma_k^z,   \nonumber \\ 
    E_j(t)=& \ E^x_j(t)\cos\left(\nu_j t\right)+E^y_j(t)\sin\left(\nu_j t\right) ,\nonumber
\end{align}
where $\nu_j$ is the frequency of the drive on the $j$-th qubit, $E^x_j(t)$ and $E^y_j(t)$ the two quadratures of the field, $d_j$ the $j$-th qubit's dipole matrix element, and $\mathcal{L}$ the driven subset. Typically, the qubit's transition energy, $\omega_j$, is much larger than the interaction, $J_{jk}$, and hence $\ket{0,0,...,0}$ is the ground state of the undriven Hamiltonian and can be initialised by cooling.

In the frame rotating with the qubits' frequencies, described by the unitary transformation $U_0(t)=e^{i\sum_{j}\frac{\omega_j}{2}\sigma^z_j t}$, the Hamiltonian in the rotating wave approximation is
\begin{align}
    H(t)&\approx \sum_{j\in \mathcal{L}} \frac{1}{2}  \left[ \Omega_j(t) \sigma^x_j+ \Omega'_j(t)\sigma^y_{j}\right]+\sum_{jk}J_{jk}\sigma_j^z \sigma_k^z ,
\end{align}
 where 
 \begin{align}
 \Omega_j(t)=\Omega^x_j(t) \cos(\delta_j t)+\Omega^y_j(t)\sin(\delta_j t),\nonumber \\
  \Omega'_j(t) = \Omega^y_j(t) \cos(\delta_j t)-\Omega^x_j(t)\sin(\delta_j t).
 \end{align}
Here,  $\Omega^{x,y}_j(t)\equiv d_j E^{x,y}_j(t)$ is the Rabi frequency and $\delta_j\equiv \nu_j-\omega_j$ is the detuning.
\subsection*{Hamiltonian decomposition into commuting blocks}
Computing the unitary evolution of a many-body Hamiltonian like $H(t)$ is in general intractable due to the exponential complexity of the wave function, unless one can decompose the Hamiltonian into a sum commuting few-qubit blocks, i.e., $H(t)=\sum_{l}H_l(t)$, where all the $H_l(t)$ are mutually commuting. The unitary evolution after a time duration $T$ is then $U(T)=\prod_l U_l(T)$ where $U_l(T)=\mathcal{T} e^{-i\int_{0}^{T}dt' H_l(t')}$ and $\mathcal{T}$ is the time-ordering operator. $U_l(T)$ can be efficiently computed since it involves only a few qubits. The $U_l(T)$ factors are also mutually commuting and can be seen as \textit{parallel} quantum gates applied on separate qubit blocks. 

\begin{figure}[b]
\centering
\includegraphics[width=\columnwidth]{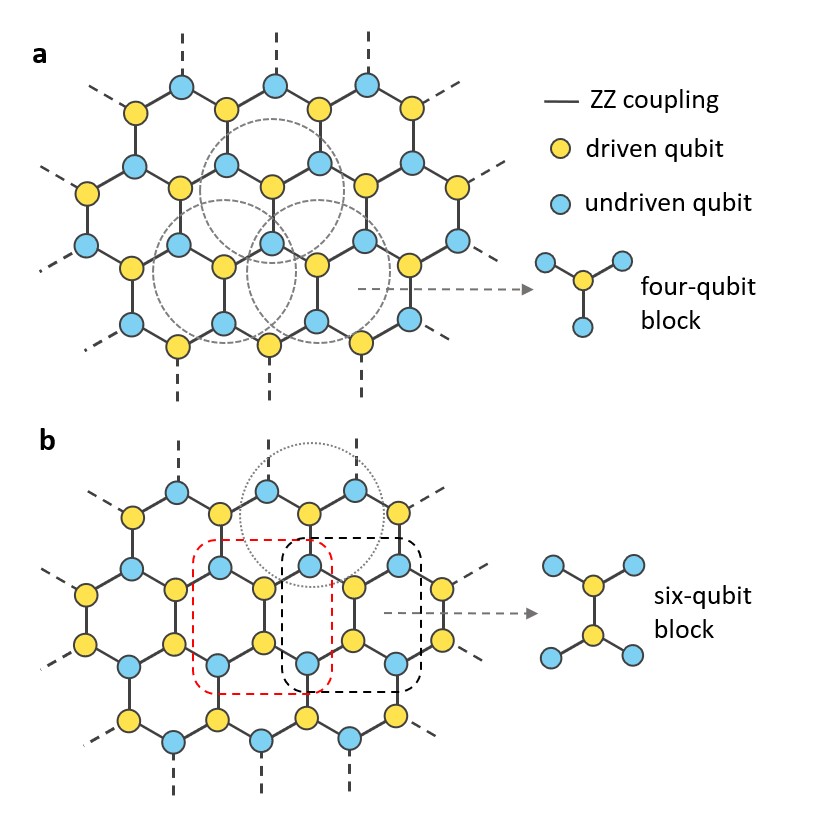}
\caption{\label{fig:2D} a) An arbitrarily large 2D array of coupled qubits in a honeycomb array. Each link represents a ZZ coupling term.  A yellow (cyan) vertex represents a driven (undriven) qubit. The total Hamiltonian can be decomposed into a sum of identical four-qubit blocks that commute with each other (enclosed in the dashed circles). An undriven qubit is shared by three neighboring blocks. (b) The pattern of driving for implementing a two-qubit gate, resulting in a central row of identical six-qubit blocks (enclosed by the dashed rectangles). The rest of the array can be decomposed into the four-qubit blocks as in (a).}
\end{figure}
We find that the simplest geometry that allows the decomposition of $H(t)$ into few-qubit commuting blocks is a honeycomb array of qubits  with nearest neighbor ZZ coupling, as shown in Fig.~\ref{fig:2D}. We consider an alternating driving pattern where only the subarray colored in yellow in Fig.~\ref{fig:2D}a is driven, then $H(t)=\sum_{j\in \mathcal{L}}H_j(t)$ where $\mathcal{L}$ is the driven subarray and 
\begin{align}\label{eq:blockH}
   H_j(t)=\frac{1}{2} \left[\Omega_{j}(t) \sigma^x_{j}+\Omega'_{j}(t) \sigma^y_{j}\right]+\sum_{k\in \text{NB}_j}J_{jk}\sigma_{j}^z \sigma_{k}^z,
\end{align}
where $\text{NB}_j$ is the set of the three nearest neighbors of qubit $j$. Each of the block Hamiltonians, $H_j$, has only four qubits and they commute with each other. This can be seen in a more transparent way by the graphical representation in Fig.~\ref{fig:2D}a. Each link in the graph represents a $ZZ$ coupling term; a yellow vertex represents a single qubit driving term, and a cyan vertex represents an undriven qubit and hence nothing in $H(t)$.  All the ZZ terms commute with each other. A yellow vertex does not commute only with the three vertices connected to it, because $\sigma^x$ and $\sigma^y$ do not commute with $\sigma^z$. Thus, the total Hamiltonian can be expressed as a sum of commuting four-qubit blocks enclosed by the dash circles in Fig.~\ref{fig:2D}a. Note that each block has only one driven qubit in the center. The qubits at the intersection of two neighboring blocks must not be driven for the commutativity to hold. We will show below that it is possible to implement a single qubit gate on the driven qubit without changing the state of the undriven qubits, at the end of the gate, despite the permanent ZZ interaction in the block. 

The driving pattern need to be modified slightly for implementing two-qubit gates. In a conventional device with tunable couplers the qubit-qubit interaction is turned on only when a two-qubit entangling gate is applied. In our case the ZZ coupling is always on, and in general it entangles all the qubits at all times. However, we find that it is still possible to implement a specific two-qubit entanging gate, for example the CNOT gate, between two targeted qubits by driving both. Turning on the drives on two neighboring qubits results in the pattern of Fig.~\ref{fig:2D}b where the central row is built from identical six-qubit blocks. The rest of the array can be driven in the alternating pattern as before. The reader may wonder why the six-qubit blocks are required for the entire central row when only one two-qubit gate is needed. This is necessary for applying the identity operators on all undriven qubits for cancelling the actions of the fixed ZZ coupling, which requires that any undriven qubit must have at least one neighboring driven qubit (more details below). If a link, $J_{jk}\sigma^z_j\sigma^z_k$, is not connected to any driven qubit, then it commutes with every other terms in the Hamiltonian, and its contribution to the total unitary evolution is simply the factor $e^{-iJ_{jk}\sigma^z_j\sigma^z_k T}$, which cannot be cancelled due to the absence of control.

\subsection*{Applying gates using optimal control}\label{sec:gate}
\begin{figure}[t]
\centering
\includegraphics[width=0.85\columnwidth]{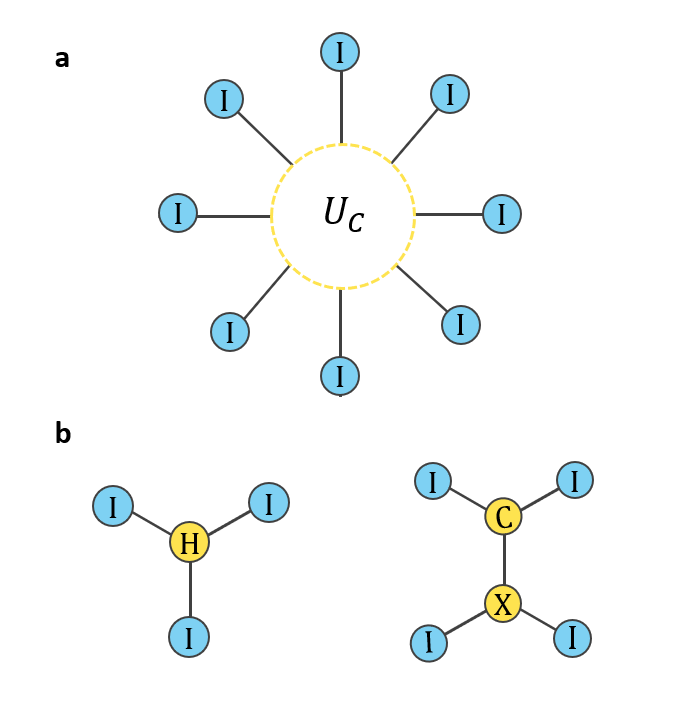}
\caption{\label{fig:block} a) A star graph with undriven qubits on the boundary and a driven subet of one or more qubits in the center. Each undriven qubit must be connected to at least one driven qubit. By utilizing optimal control on the driven subset it is possible to apply a unitary of the form $U_{\mathcal{C}}\otimes I_{\mathcal{B}}$ where $U_{\mathcal{C}}$ is a target unitary on the central driven subset, and $I_{\mathcal{B}}\equiv I\otimes I\otimes\dots\otimes I$ is the identity in the Hilbert space of the undriven qubits on the boundary (see text).  b) Examples of applying a Hamadard gate on the four-qubit block and CNOT (CX) gate on the six-qubit block of Fig.~\ref{fig:2D}. In both cases the identity operators are applied on all undriven qubits.}
\end{figure}
We now describe how to apply targeted gates on the driven qubits while at the same time apply the identity operators on the neighboring undriven qubits. Note that the four and six qubit blocks in Fig.~\ref{fig:2D}a and \ref{fig:2D}b have the form of a star graph where only a central subset of qubits is driven, as depicted in Fig.~\ref{fig:block}a. Our method lies in the key numerical finding that, for such a star graph, it is possible to use optimal control algorithm to find pulse shapes, $\Omega_j^{x,y}(t)$ where $j\in \textit{driven subset}$, to implement a unitary operation of the type $U_{\mathcal{C}}\otimes I_{\mathcal{B}}$, where $U_{\mathcal{C}}$ is a unitary acting on the driven subset, and $I_{\mathcal{B}}$ the identity matrix acting on the undriven subset at the boundary. The net effect is that the gate $U_{\mathcal{C}}$ is applied to the driven subset while the rest remains unchanged. If the driven subset has one (two) qubit, then $U_{\mathcal{C}}$ is a single qubit (two-qubit) gate.

Obviously the qubits on the boundary are acted on by the ZZ interactions, and hence their states are changed, during the pulse, but by choosing the right shape one can use the combined effect of the central driving term and the ZZ connectors to ensure that the identity operators are applied at then end of the pulse, removing the ZZ interactions in a stroboscopic fashion. This can be partly understood by looking at the Baker–Campbell–Hausdorff expansion. For the four-qubit block with the Hamiltonian of Eq.~\eqref{eq:blockH} , for example, the unitary evolution in a small time step is
\begin{align}
    e^{-i H_j(t) dt}&= e^{-i \left(H^d_j(t)+H^{int}_j\right)dt}\nonumber \\
    &\approx e^{-i H_j^d(t) dt}e^{-i H^{int}_j dt}e^{-[H^d_j(t),H^{int}_j]dt^2/2},
\end{align}
where $H^d_j(t)\equiv (1/2) \left[\Omega_{j}(t) \sigma^x_{j}+\Omega'_{j}(t) \sigma^y_{j}\right]$ is the driving term, $H^{int}_j\equiv \sum_{k\in \text{NB}_j}J_{jk}\sigma_{j}^z \sigma_{k}^z$ the ZZ interaction terms, and $[H^d_j(t),H^{int}_j]$ the commutator of the two. While the first term is responsible for applying a gate on the driven qubit, the second is the unwanted evolution due to the ZZ interaction. Since $[H^d_j(t),H^{int}_j]=\sum_{k\in \text{NB}_j}iJ_{jk}\left[\Omega_j(t)\sigma_{j}^x-\Omega'_j(t)\sigma_{j}^y\right] \sigma_{k}^z$, it is obvious that the third term allows partial control of the undriven qubits, labelled by $k$, through shaping $\Omega_j^{x,y}(t)$; and we find that this is sufficient for undoing the evolution due to the ZZ interaction. 

Using optimal control we are able to obtain pulses for realising the unitary operator $U_{\mathcal{C}}\otimes I_{\mathcal{B}}$ with maximum fidelity, $F=1$, up to numerical precision, where $U_{\mathcal{C}}$ is the Hadamard, $\pi/8$ and the \emph{direct} identity gate on the the single driven qubit of the four-qubit block (we use ``\emph{direct} identity gate" to refer to an identity gate applied on an driven qubit to differentiate it from the identity operators applied on the undriven qubits). The same was achieved where $U_{\mathcal{C}}$ is the CNOT gate and the direct two-qubit identity gate, $I\otimes I$, on the two driven qubits of the six-qubit block of Fig.~\ref{fig:block}b. Note that in all of these examples the identity operators are applied on the undriven qubits. These one and two qubit gates form a unversal set, i.e., a set from which any muti-qubit unitary can be approximated from with arbitrary precision, \emph{allowing the implementation of an arbitrary quantum circuit} \cite{Nielsen2000}. More details of the optimal control algorithm and pulse shapes are given later where we discuss the robustness of the these gates against parameter uncertainty in the Hamiltonian.

\subsection*{Implementing quantum circuits}

An implementation of quantum computation on the array is illustrated in Fig.~\ref{fig:circuit} where the driven subarray is varied from one step to the next to apply the target gate on the right qubits. As can be seen in Fig.~\ref{fig:circuit} the identity operators are applied on the undriven qubits at every step. Note that the commuting blocks are not fixed, but are changed constantly during the execution of a quantum circuit, depending on where the gates are applied and whether they are single or two-qubit gates. 

\begin{figure}[h]
\centering
\includegraphics[width=\columnwidth]{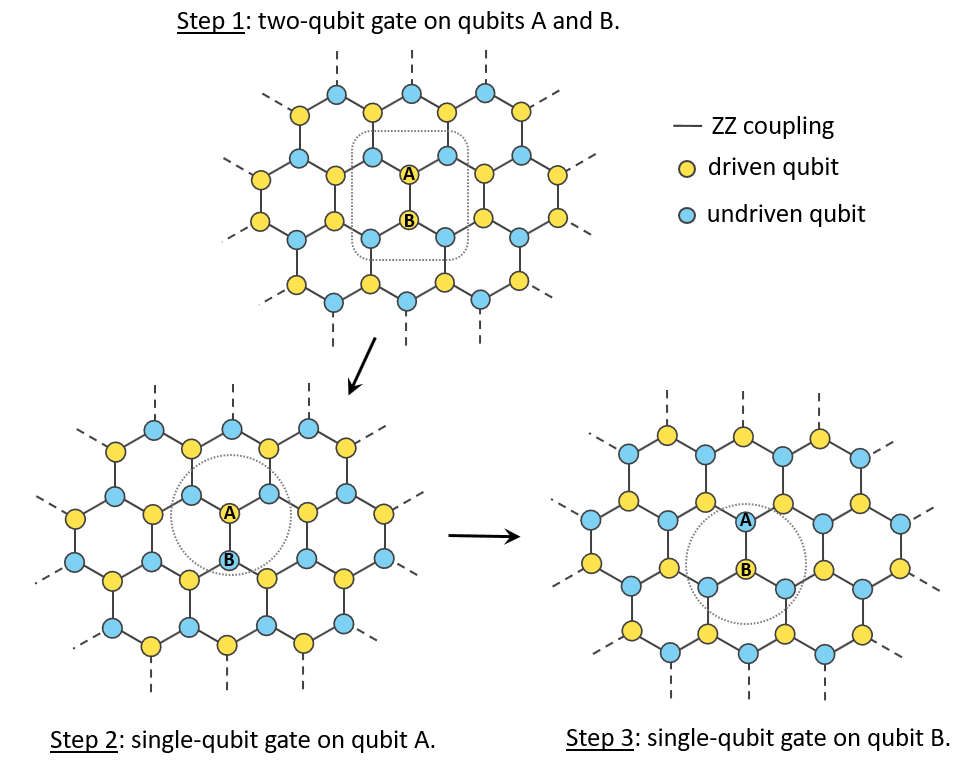}
\caption{\label{fig:driving_pattern} An illustration of how the blocks are changed/shifted during an implementation of the simple quantum circuit described in the text.}
\end{figure}

We show in Fig.~\ref{fig:driving_pattern} a simple example of how the driven/undriven subarrays and the blocks are varied during the implementation of a simple quantum circuit. Consider the following sequence of gates on two qubits, denoted by A and B: a two-qubit gate on A and B, followed by a single qubit gate on A, and then a single qubit gate on B. Note how the driven/undriven subarrays and the blocks are changed at each step. In step 1 both qubits, A and B, are driven in a six-qubit block, in step 2 only qubit A is driven in a four-qubit block and qubit B now becomes an undriven qubit, and in step 3 qubit B is driven and qubit A undriven. At each step gates can also be implemented in parallel on the driven qubits other than A and B. This parallel processing helps reduce the number of steps in a computation. If there is no gate on a driven qubit at a given step one simply applies the direct identity gate to keep its state unchanged. The undriven qubits are always subjected to the identity operators at all steps. Following this example it is straightforward to derive the driving pattern for an arbitrary quantum circuit.

In our method the undriven subarray is crucial for the Hamiltonian decomposition into commuting blocks, but this means that there are always qubits that have no gate at a given step, leading to an overhead in the number of steps compared with conventional quantum computation with tunable couplers. The exact amount of this overhead  depends on the gate configuration of a circuit. However, as shown in Fig.~\ref{fig:2D}, the number of undriven qubits is no more than half of the total qubits, and it can be shown that the overhead is at worst a factor of 2, which does not change the computational complexity of a quantum algorithm.

Now we discuss \emph{readout} for the qubit array which obviously takes a finite duration. At the end of the computation the drives are turned off and the system evolves according to the background Hamiltonian  $H_0=\sum_{j,k}J_{j,k}\sigma_{j}^z \sigma_{k}^z$. The final wavefunction can be expanded in the computational basis, $\ket{\psi_f}~=~\sum_{j_1,...,j_N = 0,1}c_{j_1,...,j_N}\ket{j_1,...,j_N}$. Since $H_0$ consists of exclusively $\sigma^z$ terms,  $\ket{j_1,...,j_N}$ is an eigenstate of $H_0$ and the evolution under $H_0$ introduces only addition phases in the coefficients $c_{j_1,...,j_N}$. Since the solution of a computation is usually encoded in the bit string of the final wavefunction, as is the case for the Shor's algorithm for factoring \cite{Shor1997} and HHS algorithm for solving linear systems \cite{Harrow2009}, the additional phase will not change the result.  
\subsection*{Robustness against parameter uncertainty}\label{sec:robustness}
\begin{table*}[ht] 
\centering
\begin{center}
\begin{tabular}{ |c|c|c|c|c|c|c|c| } 
\hline 
&&&&&&&\\[-1em]
$\Delta J/\bar{J}$ & $\Delta \alpha$  & $\Delta \delta/\bar{J}$ & $H\otimes I_\mathcal{B}$   & $\pi/8\otimes I_\mathcal{B}$ & $I\otimes  I_\mathcal{B}$ & CX$\otimes I_\mathcal{B}$ & $I_2\otimes I_\mathcal{B}$ 
\\ \hline
&&&&&&&\\[-1em]
0\%  & 0\% & 0\% & 10  & 10  & 10 & 10 & 10 \\ \hline
&&&&&&&\\[-1em]
0.1\%  & 0.1\% &0.1\% & 5.6 (4.5)  & 5.5 (4.6)  & 5.5 (4.7) & 4.4 (4.1) & 4.6 (4.3) \\ \hline
&&&&&&&\\[-1em]
1\%  & 1\%   &0.1\% & 5.6 (2.8)  & 5.5 (2.7)  & 5.6 (2.7)&3.7 (2.4) & 4.2 (2.3)\\ \hline
&&&&&&&\\[-1em]
5\%  & 5\%  &0.1\% & 5.3 (1.4)  & 5.3 (1.2)  & 5.4 (1.1)&3.2 (1.1) &4.0 (0.9)\\
\hline
\end{tabular}
 \caption{Maximum worst-case fidelities. The figures shown are the exponents of the infidelity, i.e., $-\log_{10}\left(1-\mathcal{F}_{\text{max}}\right)$, which is the number of nines in $\mathcal{F}_{\text{max}}$. The fidelities are calculated for the single qubit gates on the four-qubit block, and the two-qubit gates on the six-qubit block of Fig.~\ref{fig:2D}. Four levels of uncertainty in the qubit-qubit coupling strength and control amplitude are considered, while the uncertainty in the detuning is kept at  $0.1\%$ of the coupling strength. The maximum amplitude of the Rabi frequencies is constrained to less than $10\bar{J}$.  $I$ and $I_2$ are the direct identity gates for one and two qubits, respectively. The gate duration is $T=2\pi/\bar{J}$, divided into $M=100$  time  bins.  The figures inside the bracket are results obtained with non-robust optimization.}
\label{tab:gates}
\end{center}
\end{table*}
We now describe the optimal control algorithm for maximising the fidelity of the unitary $\mathcal{U_C}\otimes\mathcal{I_B}$ and how to make this unitary robust against parameter uncertainty in the Hamiltonian. We divide the pulse duration, $T$, into $M$ time bins of interval $\Delta t$. In each time bin the field amplitudes are kept constant. The set of the Rabi frequencies form the control vector, $\bm{c}=\{\Omega_{jn}^{\mu}: 1\leq n\leq M; \mu=x,y; j\in \mathcal{C} \}$, where $\mathcal{C}$ is the driven subset. The unitary evolution $U_{\mathcal{G}}(T)$ of the star graph $\mathcal{G}$ is then a function of $\bm{c}$. Each qubit-qubit coupling, $J_{jk}$, and detuning, $\delta_j$, is allowed to vary independently in the uncertainty intervals $[\bar{J}-\Delta J/2,\bar{J}+\Delta J/2]$, and $[\bar{\delta}-\Delta \delta/2,\bar{\delta}+\Delta \delta/2]$, respectively. The uncertainty in the Rabi frequencies can be caused by that in the dipole-matrix elements, or a slow drift in the drive  leading to changes in the field amplitudes from one experiment to the next. This can be modelled by replacing $\Omega_j^{x,y}(t)$ in $H_\mathcal{G}$ by $\alpha_j \Omega_j^{x,y}(t) $ where $\alpha_j$ is a dimensionless parameter that varies in the interval $[1-\Delta \alpha/2,1+\Delta \alpha/2]$. Now the unitary $U_{\mathcal{G}}(T)$ also depends on $J_{jk}$,  $\alpha_j$, and $\delta_j$. The qubit-qubit coupling and dipole-matrix element of a qubit have to be measured or estimated at the point of fabrication, so their values can change substantially after more qubits are added to the array due to additional interaction or experimental drift. In contrast, we find that it is possible to determine the frequency of every qubit in the completed array  with typically very precise spectroscopic measurement. In the array the resonant frequency of each qubit is shifted due to the ZZ interactions, but there exists a procedure of one and two-photon absorption measurements that can be combined to cancel these shifts and obtain the bare qubit frequency, $\omega_j$ (see Method). Thus,  we assume that the driving fields are tuned to resonance, $\bar{\delta}=0$, with residual detuning uncertainties much smaller than the average qubit-qubit interaction, $\Delta \delta_j/\bar{J}=0.1\%$, which is typical for superconducting qubits \cite{footnote}. This small detuning uncertainty should be achievable in most physical realisations of qubits owing to the high accuracy of spectroscopic measurements.

Denote the set of these uncertain parameters as $\bm{v}$, then the robust optimal control problem can be defined as a max-min optimization problem: We find an optimal control that maximizes the minimum fidelity over $\bm{v}$, 
\begin{align}\label{eq:fidelity}
\mathcal{F}_{\text{max}}= \max_{\bm{c}} \mathcal{F}(\bm{c}), \quad \mathcal{F}(\bm{c})=\min_{v\in\mathcal{V}} F(\bm{c},\bm{v}),
\end{align}
where $\mathcal{V}$ is the hypercube containing the possible values of $\bm{v}$, and $F(\bm{c},\bm{v})$ the gate fidelity based on the trace distance
\begin{align}
    F(\bm{c},\bm{v})=\left\vert\frac{1}{D}\, \text{tr}\!\left[U_{\mathcal{G}}^{\dagger}(T)\left(U_{\mathcal{C}} \otimes I_{\mathcal{B}} \right)\right] \right \vert^2,
\end{align}
where $D$ is the dimension of the Hilbert space of $\mathcal{G}$, $I_{\mathcal{B}}$ the identity matrix of the subset of undriven qubits on the boundary, and $U_{\mathcal{C}}$ the target unitary that we want to apply on the central driven subset. In numerical computation one chooses a set of sampling points, $\bm{v}_i$ , in $\mathcal{V}$, and find the minimum fidelity in this set. We found that when the uncertainties are all smaller than $5\%$ and $F(\bm{c},\bm{v})$ is larger than $99\%$  its minimum over $\bm{v}$ always lies at one of the extreme points of $\mathcal{V}$, i.e., one of the corners of the hypercube. Thus, we can redefine $\mathcal{F}(\bm{c})\equiv \min_{\bm{v}_i\in\mathcal{X}} F(\bm{c},\bm{v}_i)$ where $\mathcal{X}$ is the discreet set of the extreme points of $\mathcal{V}$. This drastically reduces the  number of sampling points where $F(\bm{c},\bm{v})$ has to be computed. There are $2^{n_u}$ extreme points in a hypercube of $n_u$ uncertain parameters. For example, the 6-qubit block in Fig.~\ref{fig:2D}b has 9 uncertain parameters, five $J$s, two $\alpha$s, and two $\delta$s, so we need to compute $2^9\equiv 512$ values of $F(\bm{c},\bm{v}_i)$ for any given $\bm{c}$. 

We develop a numerical computation for gradient-based robust optimal control that can handle systems with up to 12 qubits and multiple uncertain parameters. Starting with a random initial guess for $\bm{c}$, we use gradients to identify a step $\delta \bm{c}$ to maximize $\min_{i}\nabla_{\bm{c}} {F(\bm{c},\bm{v}_i)}.\delta \bm{c}$, the first order increment, so that $F(\bm{c},\bm{v}_i)$ is increased for all $\bm{v}_i$. In this way  $\mathcal{F}(\bm{c})$, the minimum fidelity over $\bm{v}_i$, can be increased to a value very close to one. In practice we find that one can also raise $\mathcal{F}(\bm{c})$ by maximizing the average fidelity, $\sum_{i=1}^{n_{\mathcal{X}}} F(\bm{c},\bm{v}_i)/n_{\mathcal{X}}$, with gradient ascent, which in some cases works faster than trying to increase $F(\bm{c},\bm{v}_i)$ for all $\bm{v}_i$. For the $n$-th time bin the time evolution is calculated by the mid-point rule $e^{-i H_{\mathcal{G}}(t_n- \Delta t/2)\Delta t}+O(\Delta t^3)$ which is accurate when the time step is small. The matrix exponentiation is sped up by using the Krylov subspace method on sparse matrices; and the gradient computed from a simple and efficient second-order formula \cite{deFouquieres2011}.

For the four-qubit block of Fig.~\ref{fig:2D}a, we derive optimal pulses to realize $U_{\mathcal{C}}\otimes I_{\mathcal{B}}$ where $U_\mathcal{C}$ is the Hadamard and $\pi/8$ gates \cite{Nielsen2000}. And for the six-qubit block of Fig.~\ref{fig:2D}b we want $U_{\mathcal{C}}$ to be the CNOT gate. These three gates form a universal set where any multi-qubit unitary can be approximated from with arbitrary precision \cite{Nielsen2000}, allowing universal quantum computing on the array. For a system with exclusively fixed coupling, it is also necessary to implement the identity gate that keeps all the qubits in a block unchanged despite the permanent interaction.

In Table~\ref{tab:gates} we show the robust fidelities obtained for the universal set and the identity gates at various levels of uncertainty. The gate duration is divided into $M=100$ time bins. For $1\%$ uncertainty the fidelity is higher than 99.999\% for the single qubit gates and exceeds 99.98\% for the two-qubit gates. Even if the uncertainty is as high as 5\% five-nines fidelities are still achieved for the single qubit gates, and above 99.94\% for the two-qubit gates. This can be improved by increasing the number of initial guesses, relaxing the constraints, or raising the number of control variables. The optimal pulse shape for the Hadamard gate is shown in Fig.~\ref{fig:optm}, and the pulse shapes for the other gates in Table~\ref{tab:gates} is given in the Supplemental Materials. We choose $T=2\pi/\bar{J}$ but this specific value of the duration is not essential; the same order of magnitude is achieved for the fidelities when $T$ is changed by 10\%. In order to see the effectiveness of robust optimal control we also calculate the fidelities with non-robust optimal control: We first neglect all the uncertainties and optimize the fidelity for the ideal case where $J_j=\bar{J}$, $\alpha_j=1$ and $\delta_j=0$ for all $j$, then we use the obtained optimal control, $\bm{c}_{\text{ideal}}$, to calculate the minimum fidelity in the hypercube, $\mathcal{F}=\min_{v\in\mathcal{V}} F(\bm{c}_{\text{ideal}},\bm{v})$. The results are shown in the parentheses of Table~\ref{tab:gates}. robust optimization improves the fidelities by two to three nines when the uncertanties are significantly large ($1\%$ and $5\%$).  

\begin{figure}[b]
\centering
\includegraphics[width=0.9\columnwidth]{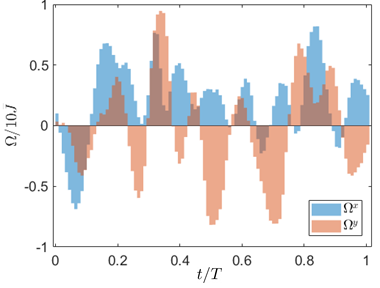}
\caption{\label{fig:optm} Optimal pulse for  the Hadamard gate at $1\%$ uncertainty in Table~\ref{tab:gates}.}
\end{figure}

\section*{Discussion}\label{sec:discuss}
A promising physical realisation of our model is the superconducting qubits including the flux and transmon qubits. A direct ZZ interaction between flux qubits can be realised by coupling the qubits inductively, as demonstrated in quantum annealers \cite{Johnson2011,Tennant2021}. There is another interesting scheme based on the inductive longitudinal coupling of the flux qubits with a common bus cavity \cite{Billangeon2015,Didier2015}, which can be scaled up to 2D arrays \cite{Richer2016}. In addition, a cross-Kerr ZZ interaction has been recently demonstrated for transmon qubits using a flux-tunable coupler \cite{Collodo2020}. While flux qubits are very good two-level systems and hence our results are applicable, for transmons the leakage to higher excited states must be accounted for in the optimal control algorithm \cite{Motzoi2009}. The ZZ coupling is also the natural interaction in a nuclear magnetic resonance quantum computer \cite{Vandersypen2001}, for which sophisticated pulse-shaping is available \cite{Yang2020}, making it a good test bed for our model.

In experimental realisations, the pulse duration, $T$, is limited by the coherence time, $T_2$, of the qubits. The  decoherence rate of a block of $N'$ qubits is enhanced by a factor of approximately $N'$ in the worst case, giving rise to a lifetime of $T_2'=T_2/N'$. The fidelity of a multi-qubit unitary on the block is then bounded by $F\leq 1-T/T_2'\equiv 1-N'T/T_2$. Therefore, to achieve a fidelity $F$ for the four-qubit and six-qubit blocks in Table~\ref{tab:gates} the pulse duration needs to be shorter than $T_2(1-F)/N'$ where $N'=4$ and 6, respectively.

Although a honeycomb array is the focus of this paper, the qubits can be arranged in any physical shape that has the same connectivity, for example a square array with each qubit connected to only three nearest neighbors. Moreover, there exist driving patterns that satisfy the conditions of commutitativity and robust control for other geometries such as square arrays and one-dimensional chains (see Supplemental Materials). One can also envisage a hybrid architecture where large clusters of fixed couplers are connected with tunable couplers, keeping the number of required tunable couplers low. Such a modular structure can help ease the technological difficulties in scaling quantum computers. 

To conclude, we find that it is feasible to implement quantum computing with accurate operations on a 2D qubit array with exclusively fixed couplers. The quantum gates are robust against significant uncertainty in the qubit's frequency, qubit-qubit and drive-qubit coupling caused by fabrication imperfection and/or slowly fluctuating fields. Both 1D and 2D geometries are possible with a robust optimization process capable of handling the minimal cluster size required for each scheme. Our proposal shows that scalability can be accelerated with  simplified hardware architecture based on fixed longitudinal coupling schemes, thus motivating further development of this coupling in different physical platforms.

\section*{Method}
\subsection*{Calculating fidelity and gradient}
We first describe how the fidelity and its gradients are calculated with the midpoint rule. The time duration is divided into $M$ equal time bins with $t_0=0$ and $t_M=T$. The field amplitudes are kept constant during each time bin. The Hamiltonian of a star graph, $\mathcal{G}$, at the midpoint of the $n-$th interval from $t_{n-1}$ to $t_{n}$ is
\begin{equation}
    H_{\mathcal{G},n}=\sum_{j\in \mathcal{C}} \alpha_j  \left[ \Omega_{jn} \sigma^x_j+ \Omega'_{jn}\sigma^y_{j}\right]+\sum_{j,k\in \mathcal{G}}J_{jk}\sigma_j^z \sigma_k^z ,\nonumber
\end{equation}
where $\alpha_j$ is the dimensionless factor introduced to model the uncertainty in the Rabi frequencies (see the Main Text), $\mathcal{C}$ is the driven subset in the center of the graph and $J_{jk}\neq 0$ only for nearest neighbors, and
\begin{align}
    \Omega_{jn}&=\Omega^x_{jn} \cos(\delta_j (t_n-\Delta t/2))+\Omega^y_{jn}\sin(\delta_j (t_n-\Delta t/2)),\nonumber \\
    \Omega'_{jn} &= \Omega^y_{jn} \cos(\delta_j (t_n-\Delta t/2))-\Omega^x_{jn}\sin(\delta_j (t_n-\Delta t/2)),\nonumber 
\end{align}
where $\Omega^{x,y}_{jn}$ are the Rabi frequencies of the driving field on the $j$-th driven qubit during the interval from $t_{n-1}$ to $t_{n}$. They are the elements of the $2MN_{\mathcal{C}}\times 1$ control vector, $\bm{c}$, where $N_{\mathcal{C}}$ is the number of qubits in the driving subset. The unitary evolution from $t_{n-1}$ to $t_{n}$, $U_n=e^{-i H_{\mathcal{G},n} \Delta t}+O(\Delta t^3)$, is computed with the $\text{expm}$ function in Matlab. For an efficient calculation of the fidelity and the gradients we compute and store all the $U_n$, and then obtain the forward and backward unitary propagation operators \cite{deFouquieres2011, Motzoi2011}, defined by
\begin{align}
    U^f_n&=U_nU_{n-1}\dots U_1, \nonumber \\
    U^b_{n+1}&=U_MU_{M-1}\dots U_{n+1}, \nonumber
\end{align}
using the recursive relations $U^f_n=U_n U^f_{n-1}$ and $U^b_{n+1}=U^b_{n+2}U_{n+1}$. Then the fidelity is
\begin{align}
     F(\bm{c},\bm{v})=\left\vert\frac{1}{2^{N_{\mathcal{G}}}}\, \text{tr}\!\left[{
     U_M^{f}}^{\dagger}\left(U_{\mathcal{C}} \otimes I_{\mathcal{B}} \right)\right] \right \vert^2, \nonumber
\end{align}
where $N_{\mathcal{G}}$ is the number of qubits in the star graph. 

For computing the gradients, we note that 
\begin{align}
  H_{\mathcal{G},n}= \sum_{\mu=x,y}\sum_{j\in \mathcal{C}} \Omega^{\mu}_{jn} K^{\mu}_{jn}+\sum_{j,k\in \mathcal{G}}J_{jk}\sigma_j^z \sigma_k^z , \nonumber
\end{align}
where
\begin{align}
    K^{x}_{jn}&=g_j\left[\sigma_j^x \cos(\delta_j (t_n-\Delta t/2)) - \sigma_j^y \sin(\delta_j (t_n-\Delta t/2))\right],  \nonumber \\
       K^{y}_{jn}&=\alpha_j\left[\sigma_j^y \cos(\delta_j (t_n-\Delta t/2)) + \sigma_j^x \sin(\delta_j (t_n-\Delta t/2))\right]. \nonumber
\end{align}
One can show that the derivative of $U_n\equiv e^{-i H_{\mathcal{G},n}\Delta t}$ with respect to $\Omega^{\mu}_{jn}$ is  \cite{deFouquieres2011}
\begin{align}
    \frac{\partial U_n}{\partial \Omega^{\mu}_{jn}}=\left\{-i \Delta t K^{\mu}_{jn} +\frac{\Delta t^2}{2}\left[H_{\mathcal{G},n},K^{\mu}_{jn} \right]\right\}U_n+O(\Delta t^3), \nonumber
\end{align}
where $\left[H_{\mathcal{G},n},K^{\mu}_{j,n} \right]$ is a commutator. It follows that the derivative of $U_M^f\equiv U^b_{n+1}U_n U^f_{n-1}$ is
\begin{align}
    \frac{\partial U_M^f}{\partial \Omega^{\mu}_{jn}}=U^b_{n+1}\left\{-i \Delta t K^{\mu}_{jn} +\frac{\Delta t^2}{2}\left[H_{\mathcal{G},n},K^{\mu}_{jn} \right]\right\}U^f_n, \nonumber
\end{align}
and from this it is straight forward to compute the gradient of the fidelity. The most computationally expensive part of the calculation is the matrix exponentiation for obtaining $U_n$, which is done $M$ times. 
\subsection*{Robust optimisation}
We use two algorithms to raise the worst-case fidelity, $\mathcal{F}(\bm{c})$, defined in Eq.~\ref{eq:fidelity}. The first is based on sequential convex programming \cite{Allen2017}. We start with initial guesses $\bm{c}_0$ and $\bm{u}_0$ for the control vector and the upper limit (trust region) of the step, respectively. Then a step $|\delta \bm{c}|<\bm{u}_0$ is found to maximize $\min_{i}\nabla_{\bm{c}}F(\bm{c},\bm{v}_i).\delta\bm{c}$, i.e., to maximize the minimum first-order increment over $\bm{v}_i$. This ensures all the fidelities at the extreme points of the hypercube, $\mathcal{V}$, are increased. The above optimization problem can be solved by sequential convex programming (SCP). We used the  YALMIP toolbox and SPDT3 package in Matlab for this purpose. If a step can be found such that $\min_{i}\nabla_{\bm{c}}F(\bm{c},\bm{v}_i).\delta\bm{c}$ is positive then we increase the upper bound $\bm{u}_0$ by 1.15, otherwise we decrease it by 2. We chose these factors as they give the fastest convergence in our numerical tests. The procedure is repeated until either the maximum iteration is reached or the step's upper bound drops below a small tolerance. The second approach is to simply maximize the average fidelity,
\begin{align}
    \bar{\mathcal{F}}(\bm{c})=\sum_{i=1}^{n_{\mathcal{X}}} F(\bm{c},\bm{v}_i)/n_{\mathcal{X}},
\end{align}
using a quasi-Newton method. Obviously this does not guarantee that the worst-case fidelity over $\bm{v}$ is increased, as the mean can be increased without increasing the minimum value in the set. However, we found that in our calculations the worst-case fidelity is always improved substantially when we maximize the average fidelity. We optimize $\bar{\mathcal{F}}(\bm{c})$ using the interior-point method implemented in Matlab's fmincon function, where the Hessian is computed from the exact gradients with the BFGS approximation. 

In our numerical tests the first algorithm is more sensitive on the initial guesses of the control parameters. For the two-qubit gates in Table~I the computation is very expensive and hence it is not practical to run the optimization with too many initial guesses. We find that for the same running time the second algorithm gives higher fidelities, and the results in Table~I are obtained with it.

\subsection*{Spectroscopic measurement of qubit frequency}
\begin{figure}[b]
\centering
\includegraphics[width=0.8\columnwidth]{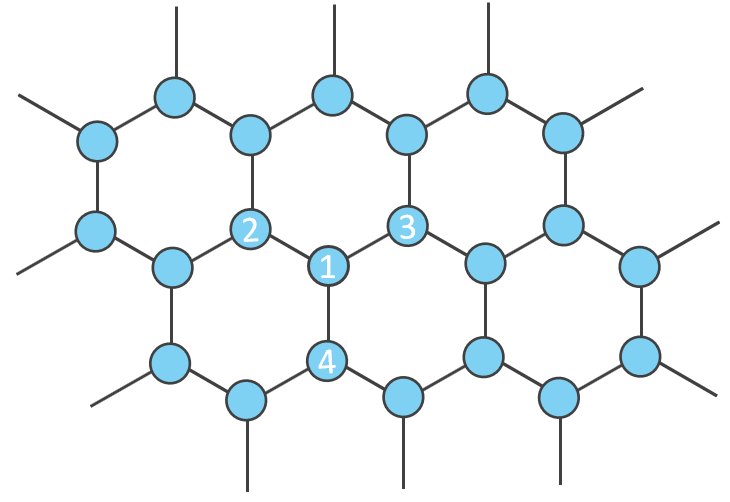}
\caption{\label{fig:measure} The frequency of qubit 1 can be determined by one and two-photon absorption measurements on the four qubits of the numbered unit cell (see text).}
\end{figure}
In this section we propose a procedure using one and two-photon absorption measurements for determining the frequency of every qubit in the 2D honeycomb array. Initially all qubits in the array are in the ground state, $\ket{0}$. We want to measure the frequency of qubit 1 in Fig.~\ref{fig:measure}. Due to the ZZ interactions with neighboring qubits causing a dispersive shift the one-photon absorption peak of qubit 1 is at
\begin{align}\label{eq:1photon1}
    \omega^p_1=\omega_1-2(J_{12}+J_{13}+J_{14}).
\end{align}
The shift is the energy cost for flipping three bonds from $\ket{00}$ to $\ket{01}$. Similarly, the one-photon absorption peaks for the other numbered qubits are
\begin{align}\label{eq:1photon2}
    \omega^p_2&=\omega_2-2 J_{21}-2\sum_{\substack{k\in \text{NB}_2 \\k\neq 1}}{J_{2k}}, \nonumber \\
    \omega^p_3&=\omega_3-2 J_{31}-2\sum_{\substack{k\in \text{NB}_3 \\k\neq 1}}J_{3k},\nonumber \\
    \omega^p_4&=\omega_4-2 J_{41}-2\sum_{\substack{k\in \text{NB}_4 \\k\neq 1}}J_{4k},
\end{align}
where $\text{NB}_j$ is the set of nearest neighbors of qubit $j$. The last sums are the dispersive shifts caused by the ZZ intreaction with neighbors that are not qubit 1. Summing all the equations in Eqs.~\eqref{eq:1photon1} and \eqref{eq:1photon2}, and using $J_{kl}=J_{lk}$, one obtains
\begin{align}\label{eq:1photon}
\sum_{j=1}^{4}\omega^p_4=\sum_{j=1}^{4}\omega_j-4\sum_{j=2}^{4}J_{1k}-2\sum_{j=2}^4\sum_{\substack{k\in \text{NB}_j \\k\neq 1}}  J_{jk}.
\end{align}
If both qubits 1 and 2 are subjected to the the same field, for example by splitting the field in a Mach–Zehnder interferometer set-up with one qubit in each arm, one can observe the two-photon transition $\ket{0}_1\ket{0}_2\rightarrow\ket{1}_1\ket{1}_2$. Now the coupling energy of the bond between qubit 1 and 2 are not changed, and the two-photon resonance happens at the frequency
\begin{align}\label{eq:2photon1}
    2\omega^p_{12}&=\omega_1+\omega_2-2 (J_{13}+J_{14})-2\sum_{\substack{k\in \text{NB}_2 \\k\neq 1}}{J_{2k}}, 
\end{align}
Similarly, the two-photon resonance frequency for the other qubit pairs are
\begin{align}\label{eq:2photon2}
   2\omega^p_{13}&=\omega_1+\omega_3-2 (J_{12}+J_{14})-2\sum_{\substack{k\in \text{NB}_3 \\k\neq 1}}{J_{3k}}, \nonumber \\
   2\omega^p_{14}&=\omega_1+\omega_4-2 (J_{12}+J_{13})-2\sum_{\substack{k\in \text{NB}_4 \\k\neq 1}}{J_{4k}}. 
\end{align}
Summing all the euqations in Eqs.~\eqref{eq:2photon1} and \eqref{eq:2photon2}, one obtains
\begin{align}\label{eq:2photon}
    2\sum_{j=2}^{4} \omega^p_{1j}=2 \omega_1 +\sum_{j=1}^4\omega_j -4\sum_{j=2}^{4}J_{1k}-2\sum_{j=2}^4\sum_{\substack{k\in \text{NB}_j \\k\neq 1}}  J_{jk}.
\end{align}
Finally, subtracting Eq.~\eqref{eq:1photon} from Eq.~\eqref{eq:2photon}, the dispersive shifts are cancelled and one arrives at
\begin{align}
    \omega_1= \sum_{j=2}^{4} 
    \omega^p_{1j}-\frac{1}{2}\sum_{j=1}^{4}\omega^p_j, \nonumber
\end{align}
which gives the bare frequency of qubit 1 in terms of the one and two-photon resonant frequencies. In most qubit platforms spectroscopic measurements are precise and thus the uncertainty in the qubit's frequency can be small.
\subsection*{Data Availability}
The data for this work are available without restriction \cite{zenodo}.
\subsection*{Code Availability}
The Matlab code for this work are available without restriction \cite{zenodo}.
\bibliography{rqcbib}

\begin{thebibliography}{26}%
\makeatletter
\providecommand \@ifxundefined [1]{%
 \@ifx{#1\undefined}
}%
\providecommand \@ifnum [1]{%
 \ifnum #1\expandafter \@firstoftwo
 \else \expandafter \@secondoftwo
 \fi
}%
\providecommand \@ifx [1]{%
 \ifx #1\expandafter \@firstoftwo
 \else \expandafter \@secondoftwo
 \fi
}%
\providecommand \natexlab [1]{#1}%
\providecommand \enquote  [1]{``#1''}%
\providecommand \bibnamefont  [1]{#1}%
\providecommand \bibfnamefont [1]{#1}%
\providecommand \citenamefont [1]{#1}%
\providecommand \href@noop [0]{\@secondoftwo}%
\providecommand \href [0]{\begingroup \@sanitize@url \@href}%
\providecommand \@href[1]{\@@startlink{#1}\@@href}%
\providecommand \@@href[1]{\endgroup#1\@@endlink}%
\providecommand \@sanitize@url [0]{\catcode `\\12\catcode `\$12\catcode
  `\&12\catcode `\#12\catcode `\^12\catcode `\_12\catcode `\%12\relax}%
\providecommand \@@startlink[1]{}%
\providecommand \@@endlink[0]{}%
\providecommand \url  [0]{\begingroup\@sanitize@url \@url }%
\providecommand \@url [1]{\endgroup\@href {#1}{\urlprefix }}%
\providecommand \urlprefix  [0]{URL }%
\providecommand \Eprint [0]{\href }%
\providecommand \doibase [0]{http://dx.doi.org/}%
\providecommand \selectlanguage [0]{\@gobble}%
\providecommand \bibinfo  [0]{\@secondoftwo}%
\providecommand \bibfield  [0]{\@secondoftwo}%
\providecommand \translation [1]{[#1]}%
\providecommand \BibitemOpen [0]{}%
\providecommand \bibitemStop [0]{}%
\providecommand \bibitemNoStop [0]{.\EOS\space}%
\providecommand \EOS [0]{\spacefactor3000\relax}%
\providecommand \BibitemShut  [1]{\csname bibitem#1\endcsname}%
\let\auto@bib@innerbib\@empty
%</preamble>
\bibitem [{\citenamefont {Arute}\ \emph {et~al.}(2019)\citenamefont {Arute}
  \emph {et~al.}}]{Arute2019}%
  \BibitemOpen
  \bibfield  {author} {\bibinfo {author} {\bibfnamefont {F.}~\bibnamefont
  {Arute}} \emph {et~al.},\ }\href {\doibase 10.1038/s41586-019-1666-5}
  {\bibfield  {journal} {\bibinfo  {journal} {Nature}\ }\textbf {\bibinfo
  {volume} {574}},\ \bibinfo {pages} {505} (\bibinfo {year}
  {2019})}\BibitemShut {NoStop}%
\bibitem [{\citenamefont {Ballance}\ \emph {et~al.}(2016)\citenamefont
  {Ballance}, \citenamefont {Harty}, \citenamefont {Linke}, \citenamefont
  {Sepiol},\ and\ \citenamefont {Lucas}}]{Ballance2016}%
  \BibitemOpen
  \bibfield  {author} {\bibinfo {author} {\bibfnamefont {C.}~\bibnamefont
  {Ballance}}, \bibinfo {author} {\bibfnamefont {T.}~\bibnamefont {Harty}},
  \bibinfo {author} {\bibfnamefont {N.}~\bibnamefont {Linke}}, \bibinfo
  {author} {\bibfnamefont {M.}~\bibnamefont {Sepiol}}, \ and\ \bibinfo {author}
  {\bibfnamefont {D.}~\bibnamefont {Lucas}},\ }\href {\doibase
  10.1103/PhysRevLett.117.060504} {\bibfield  {journal} {\bibinfo  {journal}
  {Physical Review Letters}\ }\textbf {\bibinfo {volume} {117}},\ \bibinfo
  {pages} {060504} (\bibinfo {year} {2016})}\BibitemShut {NoStop}%
\bibitem [{\citenamefont {Levine}\ \emph {et~al.}(2019)\citenamefont {Levine}
  \emph {et~al.}}]{Levine2019}%
  \BibitemOpen
  \bibfield  {author} {\bibinfo {author} {\bibfnamefont {H.}~\bibnamefont
  {Levine}} \emph {et~al.},\ }\href {\doibase 10.1103/PhysRevLett.123.170503}
  {\bibfield  {journal} {\bibinfo  {journal} {Physical Review Letters}\
  }\textbf {\bibinfo {volume} {123}},\ \bibinfo {pages} {170503} (\bibinfo
  {year} {2019})}\BibitemShut {NoStop}%
\bibitem [{\citenamefont {He}\ \emph {et~al.}(2019)\citenamefont {He},
  \citenamefont {Gorman}, \citenamefont {Keith}, \citenamefont {Kranz},
  \citenamefont {Keizer},\ and\ \citenamefont {Simmons}}]{He2019}%
  \BibitemOpen
  \bibfield  {author} {\bibinfo {author} {\bibfnamefont {Y.}~\bibnamefont
  {He}}, \bibinfo {author} {\bibfnamefont {S.~K.}\ \bibnamefont {Gorman}},
  \bibinfo {author} {\bibfnamefont {D.}~\bibnamefont {Keith}}, \bibinfo
  {author} {\bibfnamefont {L.}~\bibnamefont {Kranz}}, \bibinfo {author}
  {\bibfnamefont {J.~G.}\ \bibnamefont {Keizer}}, \ and\ \bibinfo {author}
  {\bibfnamefont {M.~Y.}\ \bibnamefont {Simmons}},\ }\href {\doibase
  10.1038/s41586-019-1381-2} {\bibfield  {journal} {\bibinfo  {journal}
  {Nature}\ }\textbf {\bibinfo {volume} {571}},\ \bibinfo {pages} {371}
  (\bibinfo {year} {2019})}\BibitemShut {NoStop}%
\bibitem [{\citenamefont {Huang}\ \emph {et~al.}(2019)\citenamefont {Huang}
  \emph {et~al.}}]{Huang2019}%
  \BibitemOpen
  \bibfield  {author} {\bibinfo {author} {\bibfnamefont {W.}~\bibnamefont
  {Huang}} \emph {et~al.},\ }\href {\doibase 10.1038/s41586-019-1197-0}
  {\bibfield  {journal} {\bibinfo  {journal} {Nature}\ }\textbf {\bibinfo
  {volume} {569}} (\bibinfo {year} {2019}),\
  10.1038/s41586-019-1197-0}\BibitemShut {NoStop}%
\bibitem [{\citenamefont {Nichol}\ \emph {et~al.}(2017)\citenamefont {Nichol},
  \citenamefont {Orona}, \citenamefont {Harvey}, \citenamefont {Fallahi},
  \citenamefont {Gardner}, \citenamefont {Manfra},\ and\ \citenamefont
  {Yacoby}}]{Nichol2017}%
  \BibitemOpen
  \bibfield  {author} {\bibinfo {author} {\bibfnamefont {J.~M.}\ \bibnamefont
  {Nichol}}, \bibinfo {author} {\bibfnamefont {L.~A.}\ \bibnamefont {Orona}},
  \bibinfo {author} {\bibfnamefont {S.~P.}\ \bibnamefont {Harvey}}, \bibinfo
  {author} {\bibfnamefont {S.}~\bibnamefont {Fallahi}}, \bibinfo {author}
  {\bibfnamefont {G.~C.}\ \bibnamefont {Gardner}}, \bibinfo {author}
  {\bibfnamefont {M.~J.}\ \bibnamefont {Manfra}}, \ and\ \bibinfo {author}
  {\bibfnamefont {A.}~\bibnamefont {Yacoby}},\ }\href {\doibase
  10.1038/s41534-016-0003-1} {\bibfield  {journal} {\bibinfo  {journal} {npj
  Quantum Information}\ }\textbf {\bibinfo {volume} {3}} (\bibinfo {year}
  {2017}),\ 10.1038/s41534-016-0003-1}\BibitemShut {NoStop}%
\bibitem [{\citenamefont {Chen}\ \emph {et~al.}(2014)\citenamefont {Chen} \emph
  {et~al.}}]{Chen2014}%
  \BibitemOpen
  \bibfield  {author} {\bibinfo {author} {\bibfnamefont {Y.}~\bibnamefont
  {Chen}} \emph {et~al.},\ }\href {\doibase 10.1103/PhysRevLett.113.220502}
  {\bibfield  {journal} {\bibinfo  {journal} {Physical Review Letters}\
  }\textbf {\bibinfo {volume} {113}},\ \bibinfo {pages} {220502} (\bibinfo
  {year} {2014})}\BibitemShut {NoStop}%
\bibitem [{\citenamefont {Fowler}\ \emph {et~al.}(2012)\citenamefont {Fowler},
  \citenamefont {Mariantoni}, \citenamefont {Martinis},\ and\ \citenamefont
  {Cleland}}]{Fowler2012}%
  \BibitemOpen
  \bibfield  {author} {\bibinfo {author} {\bibfnamefont {A.~G.}\ \bibnamefont
  {Fowler}}, \bibinfo {author} {\bibfnamefont {M.}~\bibnamefont {Mariantoni}},
  \bibinfo {author} {\bibfnamefont {J.~M.}\ \bibnamefont {Martinis}}, \ and\
  \bibinfo {author} {\bibfnamefont {A.~N.}\ \bibnamefont {Cleland}},\ }\href
  {\doibase 10.1103/PhysRevA.86.032324} {\bibfield  {journal} {\bibinfo
  {journal} {Physical Review A}\ }\textbf {\bibinfo {volume} {86}},\ \bibinfo
  {pages} {032324} (\bibinfo {year} {2012})}\BibitemShut {NoStop}%
\bibitem [{\citenamefont {Vandersypen}\ \emph {et~al.}(2001)\citenamefont
  {Vandersypen}, \citenamefont {Steffen}, \citenamefont {Breyta}, \citenamefont
  {Yannoni}, \citenamefont {Sherwood},\ and\ \citenamefont
  {Chuang}}]{Vandersypen2001}%
  \BibitemOpen
  \bibfield  {author} {\bibinfo {author} {\bibfnamefont {L.~M.~K.}\
  \bibnamefont {Vandersypen}}, \bibinfo {author} {\bibfnamefont
  {M.}~\bibnamefont {Steffen}}, \bibinfo {author} {\bibfnamefont
  {G.}~\bibnamefont {Breyta}}, \bibinfo {author} {\bibfnamefont {C.~S.}\
  \bibnamefont {Yannoni}}, \bibinfo {author} {\bibfnamefont {M.~H.}\
  \bibnamefont {Sherwood}}, \ and\ \bibinfo {author} {\bibfnamefont {I.~L.}\
  \bibnamefont {Chuang}},\ }\href {\doibase 10.1038/414883a} {\bibfield
  {journal} {\bibinfo  {journal} {Nature}\ }\textbf {\bibinfo {volume} {414}},\
  \bibinfo {pages} {883} (\bibinfo {year} {2001})}\BibitemShut {NoStop}%
\bibitem [{\citenamefont {Tsunoda}\ \emph {et~al.}(2020)\citenamefont
  {Tsunoda}, \citenamefont {Bhole}, \citenamefont {Jones}, \citenamefont
  {Jones},\ and\ \citenamefont {Leek}}]{Tsunoda2020}%
  \BibitemOpen
  \bibfield  {author} {\bibinfo {author} {\bibfnamefont {T.}~\bibnamefont
  {Tsunoda}}, \bibinfo {author} {\bibfnamefont {G.}~\bibnamefont {Bhole}},
  \bibinfo {author} {\bibfnamefont {S.~A.}\ \bibnamefont {Jones}}, \bibinfo
  {author} {\bibfnamefont {J.~A.}\ \bibnamefont {Jones}}, \ and\ \bibinfo
  {author} {\bibfnamefont {P.~J.}\ \bibnamefont {Leek}},\ }\href {\doibase
  10.1103/PhysRevA.102.032405} {\bibfield  {journal} {\bibinfo  {journal}
  {Physical Review A}\ }\textbf {\bibinfo {volume} {102}},\ \bibinfo {pages}
  {032405} (\bibinfo {year} {2020})}\BibitemShut {NoStop}%
\bibitem [{\citenamefont {Johnson}\ \emph {et~al.}(2011)\citenamefont {Johnson}
  \emph {et~al.}}]{Johnson2011}%
  \BibitemOpen
  \bibfield  {author} {\bibinfo {author} {\bibfnamefont {M.~W.}\ \bibnamefont
  {Johnson}} \emph {et~al.},\ }\href {\doibase 10.1038/nature10012} {\bibfield
  {journal} {\bibinfo  {journal} {Nature}\ }\textbf {\bibinfo {volume} {473}},\
  \bibinfo {pages} {194} (\bibinfo {year} {2011})}\BibitemShut {NoStop}%
\bibitem [{\citenamefont {Collodo}\ \emph {et~al.}(2020)\citenamefont
  {Collodo}, \citenamefont {Herrmann}, \citenamefont {Lacroix}, \citenamefont
  {Andersen}, \citenamefont {Remm}, \citenamefont {Lazar}, \citenamefont
  {Besse}, \citenamefont {Walter}, \citenamefont {Wallraff},\ and\
  \citenamefont {Eichler}}]{Collodo2020}%
  \BibitemOpen
  \bibfield  {author} {\bibinfo {author} {\bibfnamefont {M.~C.}\ \bibnamefont
  {Collodo}}, \bibinfo {author} {\bibfnamefont {J.}~\bibnamefont {Herrmann}},
  \bibinfo {author} {\bibfnamefont {N.}~\bibnamefont {Lacroix}}, \bibinfo
  {author} {\bibfnamefont {C.~K.}\ \bibnamefont {Andersen}}, \bibinfo {author}
  {\bibfnamefont {A.}~\bibnamefont {Remm}}, \bibinfo {author} {\bibfnamefont
  {S.}~\bibnamefont {Lazar}}, \bibinfo {author} {\bibfnamefont {J.-C.}\
  \bibnamefont {Besse}}, \bibinfo {author} {\bibfnamefont {T.}~\bibnamefont
  {Walter}}, \bibinfo {author} {\bibfnamefont {A.}~\bibnamefont {Wallraff}}, \
  and\ \bibinfo {author} {\bibfnamefont {C.}~\bibnamefont {Eichler}},\ }\href
  {\doibase 10.1103/PhysRevLett.125.240502} {\bibfield  {journal} {\bibinfo
  {journal} {Phys. Rev. Lett.}\ }\textbf {\bibinfo {volume} {125}},\ \bibinfo
  {pages} {240502} (\bibinfo {year} {2020})}\BibitemShut {NoStop}%
\bibitem [{\citenamefont {Nielsen}\ and\ \citenamefont
  {Chuang}(2000)}]{Nielsen2000}%
  \BibitemOpen
  \bibfield  {author} {\bibinfo {author} {\bibfnamefont {M.~A.}\ \bibnamefont
  {Nielsen}}\ and\ \bibinfo {author} {\bibfnamefont {I.~L.}\ \bibnamefont
  {Chuang}},\ }\href@noop {} {\emph {\bibinfo {title} {Quantum Computation and
  Quantum Information}}},\ Vol.~\bibinfo {volume} {1}\ (\bibinfo  {publisher}
  {Cambridge University Press},\ \bibinfo {year} {2000})\BibitemShut {NoStop}%
\bibitem [{\citenamefont {Shor}(1997)}]{Shor1997}%
  \BibitemOpen
  \bibfield  {author} {\bibinfo {author} {\bibfnamefont {P.~W.}\ \bibnamefont
  {Shor}},\ }\href {\doibase 10.1137/S0097539795293172} {\bibfield  {journal}
  {\bibinfo  {journal} {SIAM Journal on Computing}\ }\textbf {\bibinfo {volume}
  {26}},\ \bibinfo {pages} {1484} (\bibinfo {year} {1997})}\BibitemShut
  {NoStop}%
\bibitem [{\citenamefont {Harrow}\ \emph {et~al.}(2009)\citenamefont {Harrow},
  \citenamefont {Hassidim},\ and\ \citenamefont {Lloyd}}]{Harrow2009}%
  \BibitemOpen
  \bibfield  {author} {\bibinfo {author} {\bibfnamefont {A.~W.}\ \bibnamefont
  {Harrow}}, \bibinfo {author} {\bibfnamefont {A.}~\bibnamefont {Hassidim}}, \
  and\ \bibinfo {author} {\bibfnamefont {S.}~\bibnamefont {Lloyd}},\ }\href
  {\doibase 10.1103/PhysRevLett.103.150502} {\bibfield  {journal} {\bibinfo
  {journal} {Physical Review Letters}\ }\textbf {\bibinfo {volume} {103}},\
  \bibinfo {pages} {150502} (\bibinfo {year} {2009})}\BibitemShut {NoStop}%
\bibitem [{foo()}]{footnote}%
  \BibitemOpen
  \href@noop {} {}\bibinfo {note} {For superconducting qubits with $\sim 10$
  kHz frequency drift and $\sim 10$ MHz coupling, $\Delta\delta/J\sim$
  0.1\%}\BibitemShut {NoStop}%
\bibitem [{\citenamefont {de~Fouquieres}\ \emph {et~al.}(2011)\citenamefont
  {de~Fouquieres}, \citenamefont {Schirmer}, \citenamefont {Glaser},\ and\
  \citenamefont {Kuprov}}]{deFouquieres2011}%
  \BibitemOpen
  \bibfield  {author} {\bibinfo {author} {\bibfnamefont {P.}~\bibnamefont
  {de~Fouquieres}}, \bibinfo {author} {\bibfnamefont {S.}~\bibnamefont
  {Schirmer}}, \bibinfo {author} {\bibfnamefont {S.}~\bibnamefont {Glaser}}, \
  and\ \bibinfo {author} {\bibfnamefont {I.}~\bibnamefont {Kuprov}},\ }\href
  {\doibase 10.1016/j.jmr.2011.07.023} {\bibfield  {journal} {\bibinfo
  {journal} {Journal of Magnetic Resonance}\ }\textbf {\bibinfo {volume}
  {212}},\ \bibinfo {pages} {412} (\bibinfo {year} {2011})}\BibitemShut
  {NoStop}%
\bibitem [{\citenamefont {Tennant}\ \emph {et~al.}(2021)\citenamefont {Tennant}
  \emph {et~al.}}]{Tennant2021}%
  \BibitemOpen
  \bibfield  {author} {\bibinfo {author} {\bibfnamefont {D.~M.}\ \bibnamefont
  {Tennant}} \emph {et~al.},\ }\href {\doibase 10.48550/arXiv.2111.04284}
  {\enquote {\bibinfo {title} {Demonstration of long-range correlations via
  susceptibility measurements in a one-dimensional superconducting {Josephson}
  spin chain},}\ } (\bibinfo {year} {2021}),\ \bibinfo {note} {arXiv:2111.04284
  [quant-ph]}\BibitemShut {NoStop}%
\bibitem [{\citenamefont {Billangeon}\ \emph {et~al.}(2015)\citenamefont
  {Billangeon}, \citenamefont {Tsai},\ and\ \citenamefont
  {Nakamura}}]{Billangeon2015}%
  \BibitemOpen
  \bibfield  {author} {\bibinfo {author} {\bibfnamefont {P.-M.}\ \bibnamefont
  {Billangeon}}, \bibinfo {author} {\bibfnamefont {J.~S.}\ \bibnamefont
  {Tsai}}, \ and\ \bibinfo {author} {\bibfnamefont {Y.}~\bibnamefont
  {Nakamura}},\ }\href {\doibase 10.1103/PhysRevB.91.094517} {\bibfield
  {journal} {\bibinfo  {journal} {Physical Review B}\ }\textbf {\bibinfo
  {volume} {91}},\ \bibinfo {pages} {094517} (\bibinfo {year}
  {2015})}\BibitemShut {NoStop}%
\bibitem [{\citenamefont {Didier}\ \emph {et~al.}(2015)\citenamefont {Didier},
  \citenamefont {Bourassa},\ and\ \citenamefont {Blais}}]{Didier2015}%
  \BibitemOpen
  \bibfield  {author} {\bibinfo {author} {\bibfnamefont {N.}~\bibnamefont
  {Didier}}, \bibinfo {author} {\bibfnamefont {J.}~\bibnamefont {Bourassa}}, \
  and\ \bibinfo {author} {\bibfnamefont {A.}~\bibnamefont {Blais}},\ }\href
  {\doibase 10.1103/PhysRevLett.115.203601} {\bibfield  {journal} {\bibinfo
  {journal} {Physical Review Letters}\ }\textbf {\bibinfo {volume} {115}},\
  \bibinfo {pages} {203601} (\bibinfo {year} {2015})}\BibitemShut {NoStop}%
\bibitem [{\citenamefont {Richer}\ and\ \citenamefont
  {DiVincenzo}(2016)}]{Richer2016}%
  \BibitemOpen
  \bibfield  {author} {\bibinfo {author} {\bibfnamefont {S.}~\bibnamefont
  {Richer}}\ and\ \bibinfo {author} {\bibfnamefont {D.}~\bibnamefont
  {DiVincenzo}},\ }\href {\doibase 10.1103/PhysRevB.93.134501} {\bibfield
  {journal} {\bibinfo  {journal} {Physical Review B}\ }\textbf {\bibinfo
  {volume} {93}},\ \bibinfo {pages} {134501} (\bibinfo {year}
  {2016})}\BibitemShut {NoStop}%
\bibitem [{\citenamefont {Motzoi}\ \emph {et~al.}(2009)\citenamefont {Motzoi},
  \citenamefont {Gambetta}, \citenamefont {Rebentrost},\ and\ \citenamefont
  {Wilhelm}}]{Motzoi2009}%
  \BibitemOpen
  \bibfield  {author} {\bibinfo {author} {\bibfnamefont {F.}~\bibnamefont
  {Motzoi}}, \bibinfo {author} {\bibfnamefont {J.~M.}\ \bibnamefont
  {Gambetta}}, \bibinfo {author} {\bibfnamefont {P.}~\bibnamefont
  {Rebentrost}}, \ and\ \bibinfo {author} {\bibfnamefont {F.~K.}\ \bibnamefont
  {Wilhelm}},\ }\href {\doibase 10.1103/PhysRevLett.103.110501} {\bibfield
  {journal} {\bibinfo  {journal} {Physical Review Letters}\ }\textbf {\bibinfo
  {volume} {103}},\ \bibinfo {pages} {110501} (\bibinfo {year}
  {2009})}\BibitemShut {NoStop}%
\bibitem [{\citenamefont {Yang}\ \emph {et~al.}(2020)\citenamefont {Yang},
  \citenamefont {Arenz}, \citenamefont {Pelczer}, \citenamefont {Chen},
  \citenamefont {Wu}, \citenamefont {Peng},\ and\ \citenamefont
  {Rabitz}}]{Yang2020}%
  \BibitemOpen
  \bibfield  {author} {\bibinfo {author} {\bibfnamefont {X.-D.}\ \bibnamefont
  {Yang}}, \bibinfo {author} {\bibfnamefont {C.}~\bibnamefont {Arenz}},
  \bibinfo {author} {\bibfnamefont {I.}~\bibnamefont {Pelczer}}, \bibinfo
  {author} {\bibfnamefont {Q.-M.}\ \bibnamefont {Chen}}, \bibinfo {author}
  {\bibfnamefont {R.-B.}\ \bibnamefont {Wu}}, \bibinfo {author} {\bibfnamefont
  {X.}~\bibnamefont {Peng}}, \ and\ \bibinfo {author} {\bibfnamefont
  {H.}~\bibnamefont {Rabitz}},\ }\href {\doibase 10.1103/PhysRevA.102.062605}
  {\bibfield  {journal} {\bibinfo  {journal} {Physical Review A}\ }\textbf
  {\bibinfo {volume} {102}},\ \bibinfo {pages} {062605} (\bibinfo {year}
  {2020})}\BibitemShut {NoStop}%
\bibitem [{\citenamefont {Motzoi}\ \emph {et~al.}(2011)\citenamefont {Motzoi},
  \citenamefont {Gambetta}, \citenamefont {Merkel},\ and\ \citenamefont
  {Wilhelm}}]{Motzoi2011}%
  \BibitemOpen
  \bibfield  {author} {\bibinfo {author} {\bibfnamefont {F.}~\bibnamefont
  {Motzoi}}, \bibinfo {author} {\bibfnamefont {J.~M.}\ \bibnamefont
  {Gambetta}}, \bibinfo {author} {\bibfnamefont {S.~T.}\ \bibnamefont
  {Merkel}}, \ and\ \bibinfo {author} {\bibfnamefont {F.~K.}\ \bibnamefont
  {Wilhelm}},\ }\href {\doibase 10.1103/PhysRevA.84.022307} {\bibfield
  {journal} {\bibinfo  {journal} {Physical Review A}\ }\textbf {\bibinfo
  {volume} {84}},\ \bibinfo {pages} {022307} (\bibinfo {year}
  {2011})}\BibitemShut {NoStop}%
\bibitem [{\citenamefont {Allen}\ \emph {et~al.}(2017)\citenamefont {Allen},
  \citenamefont {Kosut}, \citenamefont {Joo}, \citenamefont {Leek},\ and\
  \citenamefont {Ginossar}}]{Allen2017}%
  \BibitemOpen
  \bibfield  {author} {\bibinfo {author} {\bibfnamefont {J.~L.}\ \bibnamefont
  {Allen}}, \bibinfo {author} {\bibfnamefont {R.}~\bibnamefont {Kosut}},
  \bibinfo {author} {\bibfnamefont {J.}~\bibnamefont {Joo}}, \bibinfo {author}
  {\bibfnamefont {P.}~\bibnamefont {Leek}}, \ and\ \bibinfo {author}
  {\bibfnamefont {E.}~\bibnamefont {Ginossar}},\ }\href {\doibase
  10.1103/PhysRevA.95.042325} {\bibfield  {journal} {\bibinfo  {journal}
  {Physical Review A}\ }\textbf {\bibinfo {volume} {95}},\ \bibinfo {pages}
  {042325} (\bibinfo {year} {2017})}\BibitemShut {NoStop}%
\bibitem [{zen()}]{zenodo}%
  \BibitemOpen
  \href@noop {} {}\bibinfo {note} {Data, to be uploaded to zenodo.}\BibitemShut
  {Stop}%
\end{thebibliography}%
\subsection*{Acknowledgements}
This work is supported by the UK Hub in Quantum Computing and Simulation, part of the UK National Quantum Technologies Programme with funding from UKRI EPSRC grant EP/T001062/1, and the EPSRC strategic equipment grant no. EP/L02263X/1.
\subsection*{Author contributions}
N.H.L and E.G developed the method for quantum computation. N.H.L and M.C worked on the optimisation algorithms. E.G and N.H.L designed the project.  All authors wrote the paper and discussed its implications.
\subsection*{Additional information}
\textbf{Competing interests}: The authors declare no competing interests. 
\clearpage
\onecolumngrid
\beginsupplement
\section*{Supplemental Materials}

\subsection*{Pulse shapes for the gates in Table~\ref{tab:gates}}
\begin{figure*}[b!]
\centering
\includegraphics[width=0.8\columnwidth]{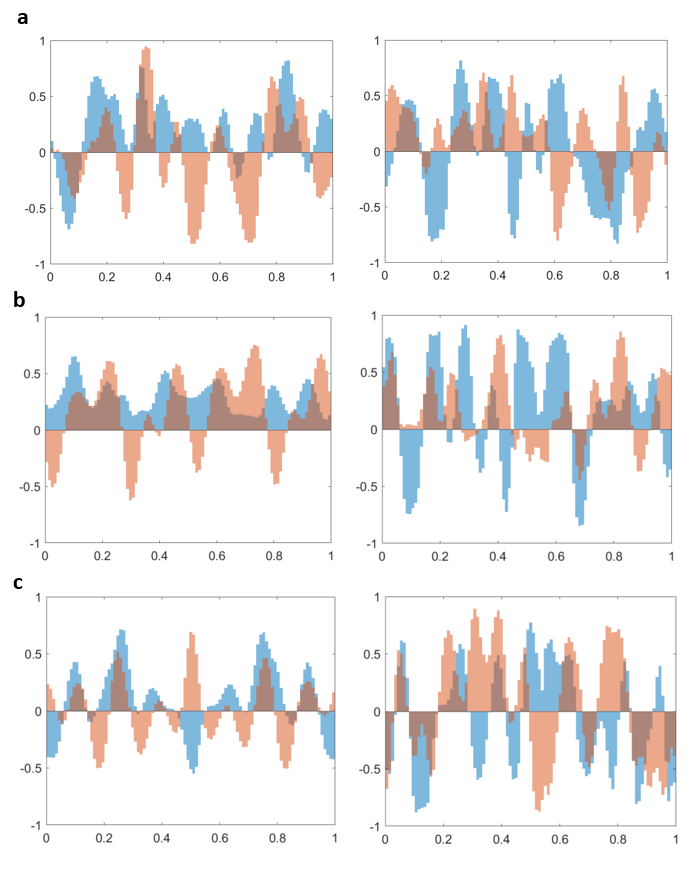}
\caption{\label{fig:1q_shape} Pulse shapes for the single qubit gates in Table I. a) Hadamard gate at 1\% uncertainty (left) and 5\% uncertainty (right) in both the control amplitude and qubit-qubit coupling. b) $\pi/8$ gate at 1\% uncertainty (left) and 5\% uncertainty (right). c) active identity gate, $\text{{\fontfamily{phv}\selectfont I
}}$, at 1\% uncertainty (left) and 5\% uncertainty (right). The detuning uncertainty  is kept at 0.1\% of the qubit-qubit coupling strength in all cases.
}
\end{figure*}

\begin{figure*}[t!]
\centering
\includegraphics[width=0.85\columnwidth]{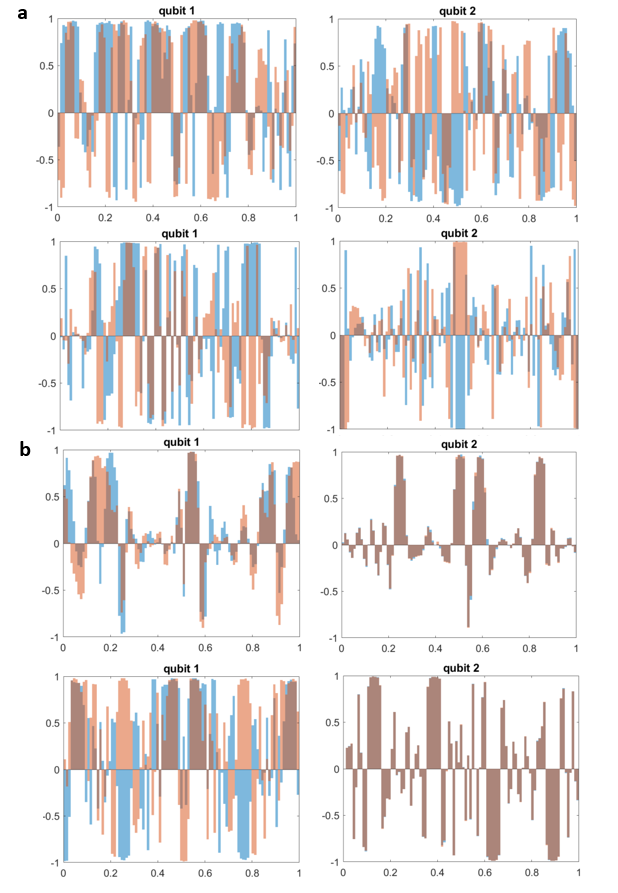}
\caption{\label{fig:2q_shape} Pulse shapes for the  two-qubit gates in Table I.  a) CNOT gate at 1\% uncertainty (top row) and 5\% uncertainty (bottom row) in both the control amplitude and qubit-qubit coupling. b) active two-qubit identity gate $\text{{\fontfamily{phv}\selectfont I
}}\otimes \text{{\fontfamily{phv}\selectfont I
}}$, at 1\% uncertainty (top row) and 5\% uncertainty (bottom row). The detuning uncertainty is kept at 0.1\% of the qubit-qubit coupling strength in all cases. Note that for qubit 1 the two quadratures of the pulses have the same shape, hence they overlap in the plot. }
\end{figure*}

In Figs.~\ref{fig:1q_shape} and \ref{fig:2q_shape} below we provide the pulse shapes for the single qubit and two-qubit gates in Table I. All the axis units and plot legends are the same as in Fig.~\ref{fig:optm}.
\clearpage 

\clearpage 

\subsection*{Commuting blocks for one dimensional chains and square arrays}

\begin{figure}[h]
\centering
\includegraphics[width=0.6\columnwidth]{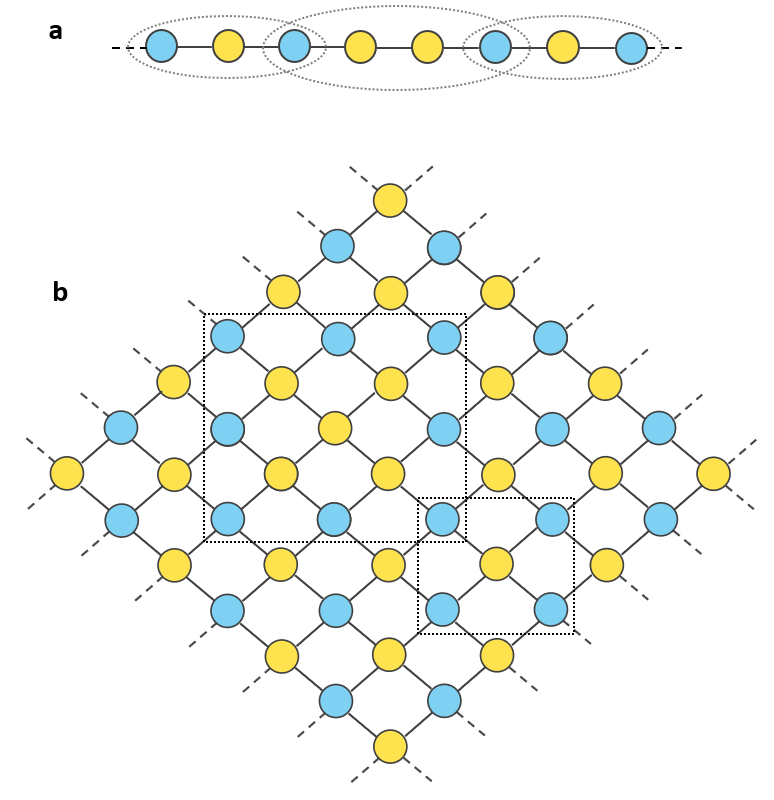}
\caption{\label{fig:1D} a) Decomposable driving pattern for a chain: One can implement any single qubit gates on the driving qubit (yellow) of the three-qubit block and any two-qubit gate between the two driving qubits of the four-qubit block. b) Decomposable driving pattern for a square array: single qubit gates can be implemented on the driving qubit of the five-qubit blocks. Two-qubit gates are more complicated in this case due to the condition for robust control (see text), requiring a thirteen-qubit block. A two-qubit gate can be implemented on any two driven qubits in the block.}
\end{figure}

In the main text we focus on the honeycomb array because it is the 2D graph with the smallest number of nearest neighbors (3), resulting in smallest possible building blocks in 2D. Here we describe how to drive 1D chains and 2D square arrays so that the Hamiltonian can be decomposed into small commuting building blocks. The result is illustrated in the graphs of Fig.~\ref{fig:1D} where the yellow vertices are driven qubits and cyan vertices undriven ones. For the square array the building block for implementing a target two-qubit gate is thirteen-qubit block where the two-qubit gate can be applied to any two nearest neighbors in the five-qubit driven subset. We could not find a simpler block due to the requirements that both qubits subjected to the two-qubit gate have to be driven, and that for robust control every ZZ link has to be connected to at least one driven qubit. We found that any driving pattern with a smaller block results in at least one link which is not connected to any driven qubit. 

\end{document}